\documentclass[journal,onecolumn]{IEEEtran}
\usepackage{multirow}
\usepackage{longtable}
\usepackage{tabularx}
\usepackage{graphicx}
\usepackage[table,xcdraw]{xcolor}
\usepackage[bottom]{footmisc}
\usepackage{empheq}
\usepackage{float}
\usepackage{cite}
\usepackage{bm}
\usepackage{amsmath,amssymb,amsfonts}
\usepackage{graphicx}
\usepackage{textcomp}
\usepackage{xcolor}
\usepackage{multirow}
\usepackage{booktabs}
\usepackage{amsfonts}
\usepackage{amsmath}
\def\BibTeX{{\rm B\kern-.05em{\sc i\kern-.025em b}\kern-.08em
    T\kern-.1667em\lower.7ex\hbox{E}\kern-.125emX}}
\usepackage{mwe}
\usepackage[utf8]{inputenc}
\usepackage[english]{babel}
\usepackage[linesnumbered,ruled,vlined]{algorithm2e}

\SetCommentSty{mycommfont}
\SetKwInput{KwInput}{Input}              
\SetKwInput{KwOutput}{Output}           
\usepackage[noend]{algpseudocode}
\usepackage[font=small,format=hang,parskip=0pt]{caption}
\usepackage[font=small,skip=0pt]{caption}
\usepackage{balance}
\usepackage{url}
\usepackage[normalem]{ulem}
\usepackage{caption}
\usepackage{subcaption}
\usepackage{xspace}
\usepackage[colorlinks=false, pdfborder={0 0 0}, breaklinks]{hyperref}
\begin{document}
\title{Towards Low-Latency and Energy-Efficient\\ Hybrid P2P-CDN Live Video Streaming}
\author{Reza~Farahani, Christian~Timmerer, and~Hermann~Hellwagner\\
\IEEEauthorblockA{\textit{Institute of Information Technology, Alpen-Adria-Universität Klagenfurt, Austria}}} 
\newcommand{\etal}{\emph{et al.\xspace}}
\newcommand{\ie}{\emph{i.e.}, }
\newcommand{\eg}{\emph{e.g.}, }
\newcommand{\etc}{\emph{etc.\xspace}}
\newcommand{\HAS}{\emph{HTTP Adaptive Streaming }}
\newcommand{\CDN}{\emph{Content Delivery Network }}
\newcommand{\DASH}{\emph{Dynamic Adaptive Streaming over HTTP }}
\newcommand{\PP}{\emph{Peer-to-Peer }}
\newcommand{\NFV}{\emph{Network Function Virtualization }}
\newcommand{\Edge}{\emph{edge computing}}
\newcommand{\QoE}{\emph{Quality of Experince}}
\newcommand{\SDN}{\emph{Software-Defined Networking}}
\maketitle
\thispagestyle{empty}
\begin{abstract}
\noindent Streaming segmented videos over the Hypertext Transfer Protocol (HTTP) is an increasingly popular approach in both live and video-on-demand (VoD) applications. However, designing a scalable and adaptable framework that reduces servers' energy consumption and supports low latency and high quality services, particularly for live video streaming scenarios, is still challenging for Over-The-Top (OTT) service providers. To address such challenges, this paper introduces a new hybrid P2P-CDN framework that leverages new networking and computing paradigms, \ie \NFV (NFV) and \Edge~for live video streaming. The proposed framework introduces a multi-layer architecture and a tree of possible actions therein (an action tree), taking into account all available resources from peers, edge, and CDN servers to efficiently distribute video fetching and transcoding tasks across a hybrid P2P-CDN network, consequently enhancing the users' latency and video quality. We also discuss our testbed designed to validate the framework and compare it with baseline methods. The experimental results indicate that the proposed framework improves user Quality of Experience (QoE), reduces client serving latency, and improves edge server energy consumption compared to baseline approaches.
\end{abstract}
\section{Introduction}
\subsection{Motivation}\label{motivation}
The rise of innovative video streaming technologies, the evolution of networking paradigms, and the growing population of users who prefer to stream online video content instead of traditional TV have collectively established video as the predominant traffic on the Internet~\cite{Sandvine}. According to the Cisco Annual Internet Report, video traffic will make up more than 60\% of the entire IP network traffic by 2023, where live video streaming has experienced remarkable popularity among all types of video traffic, constituting approximately 17\% of the total video traffic~\cite{cisco2018cisco}. \HAS (HAS) delivery systems, such as those based on the MPEG \DASH (DASH)~\cite{DASH} standard or Apple \textit{HTTP Live Streaming} (HLS)~\cite{HLS}, have emerged as the dominant technologies utilized by OTT service providers like Facebook, YouTube, and Twitch for delivering live video streaming~\cite{bentaleb2018}. In HAS-based streaming, a video is divided into small segments of fixed duration. Each segment then is encoded at multiple qualities or bitrates, known as representations. HAS clients then employ an adaptive bitrate algorithm (ABR), which enables them to adaptively download representations from media servers (\ie \CDN (CDN) servers) considering bandwidth and playout buffer conditions.
Although integrating CDN services alongside HAS-based systems has represented a significant advancement in video delivery, the substantial surge in demand for high-quality and low-latency live video poses numerous challenges for OTT providers. For instance, increasing the number of video users in popular events such as football matches can overload CDN servers. This situation may cause OTT services to deliver unsatisfactory quality and latency to end users. Consequently, OTT providers put significant pressure on backhaul networks as a bottleneck of video delivery systems~\cite{barakabitze2019qoe}.

Recent studies have demonstrated that incorporating clients' capabilities within a \textit{Peer-to-Peer} (P2P) network to establish hybrid P2P-CDN video delivery systems effectively tackles the aforementioned challenges while offering several benefits. These advantages include mitigating network congestion, enhancing streaming stability, and lowering overall delivery costs~\cite{dao2022contemporary,CDNSDNSupport2021}. Considering these benefits, many companies, such as \textit{Peer5} and \textit{Livepeer}, have adopted the utilization of peer-assisted networks, employing promising networking protocols like WebRTC to accomplish the aforementioned goals. 
Anjum~\etal~\cite{anjum2017survey} indicate that current hybrid P2P-CDN live streaming systems do not use the full potential of peers to deliver high-quality and low-latency live streaming. Hence, the primary motivation of our research is to develop a hybrid P2P-CDN live streaming system that accomplishes the following key goals: \textit{(i)} use both the computing and bandwidth capabilities offered by the P2P network, \textit{(ii)} utilize modern networking paradigms such as NFV and edge computing to propose a virtual P2P-CDN tracker server at the edge of the network, and \textit{(iii)} ensure the satisfaction of HAS client requests with improved QoE and modest latency.  
\subsection{Related Work}\label{related}
The Internet Engineering Task Force (IETF) introduced a peer assistance standard called \textit{Application-Layer Traffic Optimization} (ALTO)~\cite{alto}. ALTO enables P2P applications to receive abstract maps of network information, such as information from CDNs. Having such information allows ALTO to optimize the utilization of network resources and efficiently deliver traffic while maintaining optimal application performance. However, the ALTO protocol still has drawbacks, and its usage is limited~\cite{ellouze2013bidirectional}. As another well-known peer assistance product, Akamai proposed the NetSession Interface~\cite{zhao2013peer} that supports peer and CDN cooperation. However, it forces users to install extra software on their devices. Ma~\etal~\cite{ma2021locality} introduced machine learning-based approaches for hybrid P2P-CDN systems enabling their trackers for peer selection. However, their system does not employ edge-supported methods nor use peers' computational resources. Multiple works like~\cite{ha2017design,yousef2020enabling} customized HAS players with a prefetching module to propose such a hybrid system to reduce CDN bandwidth usage and transmission costs. Our previous works~\cite{Farahani2021eshas, Farahani2021csdn, Farahani2022leader, farahani2022ararat, farahanisarena} proposed edge- and SDN- (\SDN) assisted video streaming frameworks without considering P2P capability and mainly focusing on VoD scenarios. In other works~\cite{richter,farahani2022hybrid}, we proposed a hybrid P2P-CDN architecture for low latency live video streaming without implementation or evaluation of the energy consumption of the tracker server. However, the framework introduced in this paper is different from all the aforementioned approaches since it: \textit{(i)} uses all feasible resources of peers, \ie storage, bandwidth, and computation, \textit{(ii)} introduces a virtualized tracker server (VTS), deployable on any networking appliance, \textit{(iii)} is compatible with any ABR algorithm due to not forcing clients to modify their codes or installing additional software, and \textit{(iv)} investigates the energy consumption at the edge.
\subsection{Contributions}
To tackle these challenges mentioned in the previous sections, in this paper, we leverage HAS, P2P, CDN, NFV, and edge computing technologies to present a new hybrid P2P-CDN framework for live video streaming. Our primary objective is to enhance the serving latency and QoE for HAS clients, taking into account different resource limitations. To do that, we design an \textit{action tree} including all possible actions for serving clients' requests employed by \textit{Virtual Tracker Servers} (VTSs) at the edge of a P2P-CDN network. To validate the practical deployment of our solution, we implement the proposed approach and analyze its performance through experiments conducted in a cloud-based testbed, including 350 DASH clients. The experimental results demonstrate high users' QoE, low latency, and low edge server energy consumption compared with selected baseline approaches.
\begin{figure*}[t]
	\centering
	\includegraphics[width=.85\textwidth]{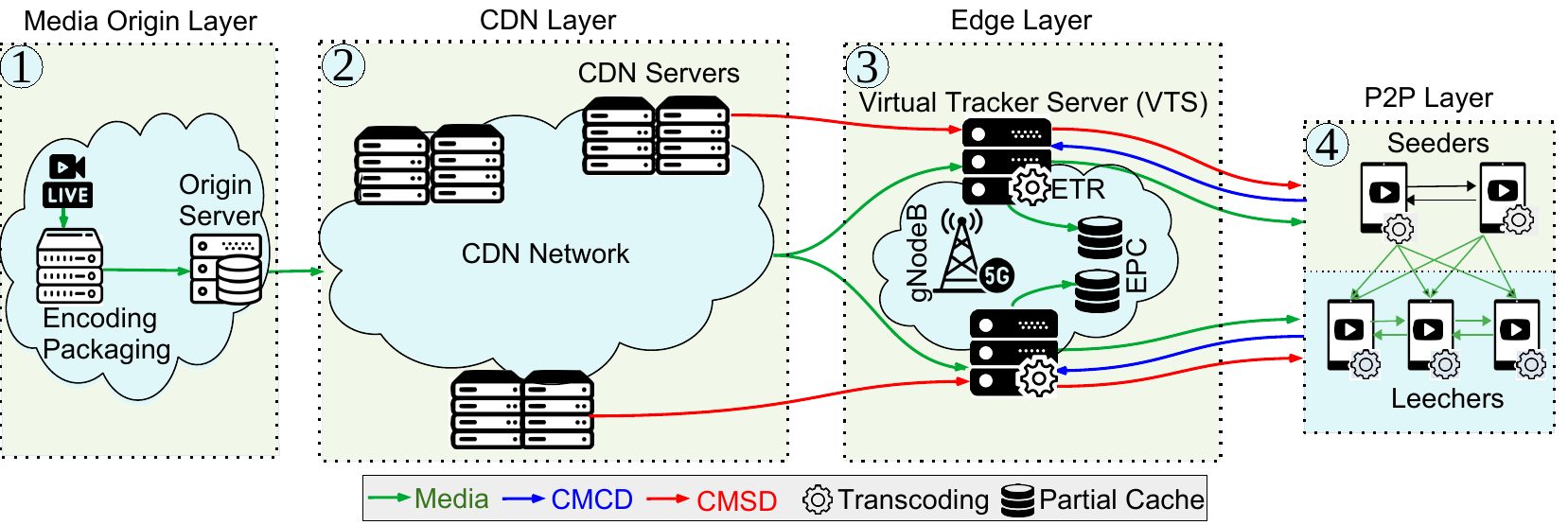}
	\vspace{.5cm}
        \caption{\small Proposed multi-layer architecture.}
	\label{arch}
\end{figure*}
\section{System Design}\label{Design}
The proposed architecture, including four core layers, is illustrated in Fig.~\ref{arch}.\\
\begin{enumerate}
    \item \textbf{Media Origin Layer:} This layer first encodes the raw live video sequences, then packages them into DASH format before storing them on the origin server. Note that this layer has the ability to package the encoded video in other formats, such as HLS or \textit{Common Media Application Format} (CMAF).\\
    \item \textbf{CDN Layer:} This layer includes a group of CDN servers, which can be OTT servers or a purchased service from CDN providers. Each server stores various parts of video sequences. Moreover, CDN servers employ \textit{Common Media Server Data} (CMSD)~\cite{CTA,CTA-cmsd} messages to periodically inform the edge layer about their cache occupancy. \\
    \item \textbf{Edge Layer:} This layer includes virtualized edge components called \emph{Virtual Tracker Servers} (VTSs), which are placed close to base stations (\eg gNodeB in 5G). VTS servers are equipped with \textit{partial cache} and \textit{video transcoding} functions to serve client requests directly from cached qualities or to construct requested qualities from existing higher qualities, respectively. In the proposed system, clients' requests are directed to a VTS server; the VTS then considers information received from the clients as CMCD messages and servers' CMSD messages, besides other monitored information, such as available bandwidth, peers' computational and power resources, and peers' joining/leaving times. It finally employs an \textit{action tree} (Fig.~\ref{tree}) to decide \textit{from where} (\ie adjacent peers, VTS, CDN servers, or origin server) and \textit{using which approach} (\ie fetch or transcode) to respond to the requested quality level. The designed action tree proposes the following potential actions that can be employed during the decision-making process (action numbering as in the figure):\\
\begin{figure*}[!t]
	\centering
	\includegraphics[width=.55\textwidth]{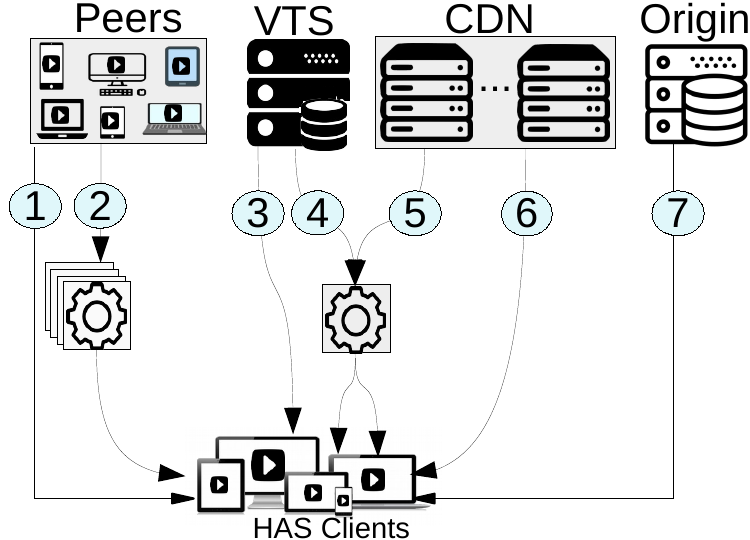}
        \vspace{.5cm}	
        \caption{\small Proposed action tree.}
	\label{tree}
\end{figure*}
    $(1)$ Use the P2P network and transmit the requested quality directly from an adjacent peer with maximum stability (\ie the least recent joining time).\\ 
    $(2)$ Transcode the requested quality from a higher quality at the most stable adjacent peer and transmit it through the P2P network.\\ 
    $(3)$ Fetch the requested quality directly from the VTS server.\\ 
    $(4)$ Transcode the requested quality from a higher quality at the VTS. \\
    $(5)$ Fetch a higher quality from a CDN server with maximum available bandwidth and transcode it at the VTS.\\
    $(6)$ Fetch the requested quality directly from a CDN server with maximum available bandwidth.\\
    $(7)$ Fetch the requested quality from the origin server. \\
    
\item \textbf{P2P Layer:} This layer utilizes all feasible peers' idle resources, \ie bandwidth, storage, and computation, to offer services like ``distributed video transcoding''. 
The P2P network is based on the \textit{tree-mesh} structure and includes two types of peers: \textit{Seeders} and \textit{Leechers}. In this scheme, seeders' requests can be served by all nodes (\ie CDNs, origin, edge, or other seeders) except leechers, while all nodes can serve leechers' requests. In the proposed system, peers periodically communicate their cache occupancies to the edge layer using \textit{Common Media Client Data} (CMCD)~\cite{CTA} messages and, in turn, receive updates from the edge layer through CMSD messages.
\end{enumerate}
\section{Evaluation Setup and Results}
\subsection{Evaluation Setup}\label{setup}
To evaluate the performance of the proposed system in a realistic large-scale environment, we select InternetMCI~\cite{zoo} as a real backbone network topology and instantiate a cloud-based testbed on the CloudLab~\cite{ricci2014introducing} environment. This testbed contains 375 elements, including  350 \textit{AStream}~\cite{AStream} DASH players (seven groups of 50 peers), running \textit{SQUAD}~\cite{wang2016squad} and \textit{BOLA}~\cite{spiteri2016bola} as hybrid and buffer-based ABR algorithms, respectively. Five Apache HTTP servers (\ie four CDN servers and an origin server), 19 OpenFlow (OF) backbone switches, 45 backbone layer-2 links, and a VTS server are used in our testbed. FFmpeg and FFmpegKit4 are used to measure the segment transcoding time on the VTS and peers, respectively. Moreover, iPhone 11 (Apple A13 Bionic, iOS 15.3), a Xiaomi Mi11 (Snapdragon 888, Android 11), and a PC (Apple M1, MacOS 12.0.1) are used to measure the transcoding time on the heterogeneous P2P network. Power consumption is measured by device tools, \eg \textit{Android Energy Profiler} and \textit{Android Battery Manager}. The \textit{CodeCarbon}~\cite{codecarbon} project measures power consumption for PC-type clients. Furthermore, we use a standard QoE model~\cite{p1203} to evaluate the system's performance regarding objective QoE. Table~\ref{tab:setup} shows the details of other configuration parameters.
\begin{table}[!t]
\centering
\caption{\small{Experimental setup.}}
\label{tab:setup}
\begin{tabular}{|l|l|}
\hline
Cache Replacement Policy                     & \multicolumn{1}{c|}{Least Recently Used (LRU)}\\ \hline
Maximum Size of CDN Caches                   & \multicolumn{1}{c|}{40\% of the video dataset}\\ \hline
Maximum Size of VTS Caches                   & \multicolumn{1}{c|}{5\% of the video dataset}\\ \hline
Maximum Size of Peer Caches                  & \multicolumn{1}{c|}{5 segments}\\ \hline
Number of Live Channels                         & \multicolumn{1}{c|}{5 channels}\\ \hline
Segment Duration                             & \multicolumn{1}{c|}{2 sec}\\ \hline
Bitrate Ladder~\cite{lederer2012dynamic}     & \multicolumn{1}{c|}{\{(0.089,320p), (0.262,480p), (0.791,720p), (2.4,1080p), (4.2,1080p)\}{[}Mbps, content resolution{]}} \\ \hline
Links' Bandwidth (CDNs to VTS)               & \multicolumn{1}{c|}{100 Mbps}\\ \hline
Links' Bandwidth (Origin to VTS)             & \multicolumn{1}{c|}{50 Mbps}\\ \hline
Channel Access Probability Model             & \multicolumn{1}{c|}{Zipf ($\alpha$ = 0.7)}\\ \hline
Monitoring Interval                          & \multicolumn{1}{c|}{1 sec}\\ \hline
\end{tabular}
\end{table}
\subsection{Evaluation Results}\label{result}
Executing computationally-intensive tasks, such as video transcoding, on peers must be fast enough to avoid imposing significant delays on the system and should not excessively drain the peers' batteries. Otherwise, clients' requests may be served by other actions, leading to network and edge server congestion. Thus, the first experiment aims to determine the power consumption (\ie power values are given as kWh$\times10^{-3}$ and percentage of battery usage) and the percentage of battery usage when various operations are concurrently executed on a single peer, such as playing video I and transcoding video II concurrently for a five-minute video (equivalent to 150 segments). The results, as presented in Table~\ref{tab:powerconsumption}, reveal that the transcoding (TR) operation consistently consumes the most power while playing back a video (PLY) incurs significantly lower battery usage. Furthermore, combining the PLY and TR tasks does not impose a substantial burden on the peers' batteries when compared to the energy consumed by individually upscaling a video.
\begin{table}[t]
\centering
\caption{\small Power consumption and battery usage of different operations for a 5-min.~video on peers.}
\resizebox{0.6\linewidth}{!}{%
\label{tab:powerconsumption}
\scriptsize
        \begin{tabular}{l|cc|cc}\toprule
        \multirow{2}{*}{\textbf{Operation}} &\multicolumn{2}{c|}{\textbf{TR: 791k$\rightarrow$262k}} &\multicolumn{2}{c}{\textbf{TR: 4219k$\rightarrow$2484k}} \\\cmidrule{2-5}
        &\textbf{Power (kWh$\times10^{-3}$)} &\textbf{Battery} &\textbf{Power (kWh$\times10^{-3}$)} &\textbf{Battery} \\\midrule
        PLY &119 &0.39\% &152 &0.49\% \\
        TR &130 &0.42\% &352 &1.14\% \\
        TR + PLY &237 &0.77\% &479 &1.55\% \\
        \bottomrule
        \end{tabular}}
\end{table}
\begin{table}[t]
\centering
\caption{\small Transcoding times and VMAF scores for a 3-min.~video on different peers.}
\resizebox{0.65\linewidth}{!}{%
\label{tab:transcodingTimeVMAF}
\scriptsize
    \begin{tabular}{cc|cc|cc}\toprule
    \centering
    \multirow{2}{*}{\textbf{Input BR (bps)}} &\multirow{2}{*}{\textbf{Target BR (bps)}} &\multicolumn{2}{c|}{\textbf{Client PC}} &\multicolumn{2}{c}{\textbf{Client Mobile}} \\\cmidrule{3-6}
    & &\textbf{Time (s)} &\textbf{VMAF} &\textbf{Time (s)} &\textbf{VMAF} \\\midrule
    4219k &89k &4.31 &15.38 &15.98 &13.75 \\
    4219k &262k &5.33 &44.61 &18.32 &42.13 \\
    4219k &791k &11.74 &76.21 &39.28 &73.14 \\
    4219k &2484k &20.44 &93.33 &74.91 &91.53 \\\midrule
    2484k &89k &3.80 &14.35 &16.55 &13.01 \\
    2484k &262k &4.83 &42.27 &18.82 &40.02 \\
    2484k &791k &11.36 &71.56 &39.76 &69.06 \\\midrule
    791k &89k &2.05 &12.21 &10.43 &11.24 \\
    791k &262k &3.35 &36.33 &14.81 &34.76 \\\midrule
    262k &89k &1.28 &11.01 &5.85 &10.32 \\
    \bottomrule
    \end{tabular}}
\end{table}

In the next experiment, we assess the transcoding times on peers and explore the impact of peer transcoding on users' perceptual quality. We employ the \textit{Video Multi-method Assessment Fusion} (VMAF)~\cite{VMAF} scores to evaluate the perceptual quality for a three-minute video. As depicted in Table~\ref{tab:transcodingTimeVMAF}, the transcoding process for the entire video takes 1.28 to 20.44 seconds (0.014 to 0.22 seconds per segment) on PC peers and 5.85 to 74.91 seconds (0.065 to 0.8 seconds per segment) on mobile peers. 

\begin{figure*}[!t]
	\centering
	\includegraphics[width=1\textwidth]{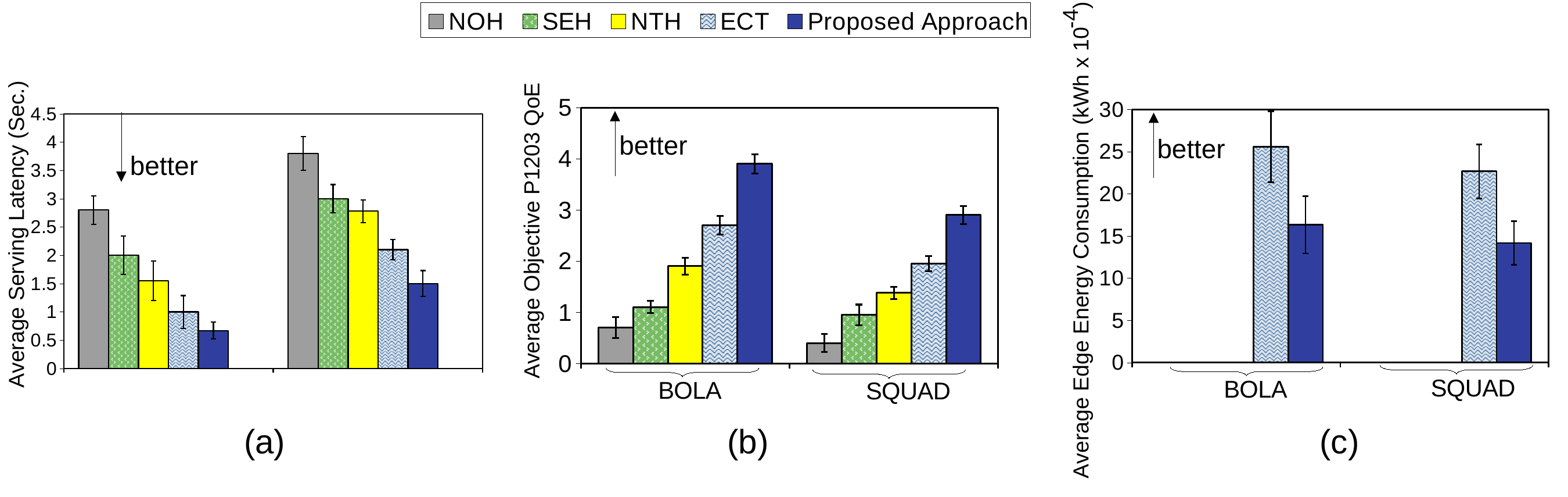}
	\caption{\small Performance of the proposed framework compared with baseline methods in terms of (a) average serving latency, (b) average objective P1203 QoE, (c) average edge server energy consumption for 350 clients running the BOLA and SQUAD ABR algorithms.}
	\label{results}
\end{figure*}

In the final experiment, we run testbed experiments to analyze the performance of the proposed framework compared to the following baseline methods:
\begin{enumerate}
    \item \textbf{Non Hybrid (NOH)}: The NOH system is a regular CDN-based streaming approach that does not incorporate any P2P support.
    \item \textbf{Simple Edge-enabled Hybrid (SEH)}: The SEH system utilizes a basic VTS server without caching and transcoding capabilities. In this system, peers can only be served through one of the actions 1, 6, or 7 (Fig.~\ref{tree}).
    \item \textbf{Non Transcoding-enabled Hybrid (NTH)}: As common in similar works, the NTH-based system has caching capability but lacks transcoding capability. In this approach, peers are only served through one of the actions 1, 3, 6, or 7 (Fig.~\ref{tree}). 
    \item \textbf{Edge Caching/Transcoding Hybrid (ECT)}: An ECT-enabled system does not involve transcoding at the peer side. In this approach, requests can be served through all actions except action 2. 
\end{enumerate}
We first calculate the average serving latency (in seconds) for all clients, encompassing both transmission latency and computational latency, as an integral component of the end-to-end (E2E) delay. As illustrated in Fig.~\ref{results}(a), the proposed system exhibits the best performance for both ABR algorithms concerning this metric. By efficiently downloading requested segments through the most suitable actions with shortened serving time (combining computation and transmission), and from nodes with available resources in terms of bandwidth and computation, the serving time is significantly reduced. 

While improving latency is crucial, it is equally important to have a comprehensive standard model to analyze the system's behavior concerning the enhancement of users' QoE. For this purpose, we employ the standard objective QoE model P.1203~\cite{p1203}. As depicted in Fig.~\ref{results}(b), the proposed system exhibits superior performance for both ABR algorithms in this aspect as well. 
We further analyze the system's performance in terms of edge energy consumption (measured in kWh) required for running edge transcoding. As evident from Fig~\ref{results}(c), the proposed system outperforms the ECT system (the only transcoding-enabled baseline system) and shows reduced energy consumption. This is primarily because of serving requested representations through \textit{(i)} distributed caching, \ie the representations cached in all layers of the network (CDN, P2P, and edge), and \textit{(ii)} distributed transcoding, the representations transcoded by transcoding-enabled peers, which allows our system to transcode fewer segments at the edge.

\section{Conclusion}
This paper presented a novel hybrid P2P-CDN live video delivery system, incorporating modern networking paradigms such as NFV and edge computing, along with distributed video transcoding. The system introduced a multi-layer architecture and an action tree, encompassing all possible actions to serve HAS clients from different nodes (\ie peers, CDNs, edge, origin server), ensuring satisfactory users' QoE and acceptable latency. To evaluate the system's performance, a cloud-based testbed with 350 clients was established, and multiple experiments were conducted. The experimental findings confirmed the superiority of the proposed approach in terms of users' QoE, serving latency, and edge server's energy consumption compared to competitive schemes.
\bibliographystyle{main}

\vspace{-1cm}
\begin{IEEEbiography} [{\includegraphics[width=1in,height=1.25in,clip,keepaspectratio]{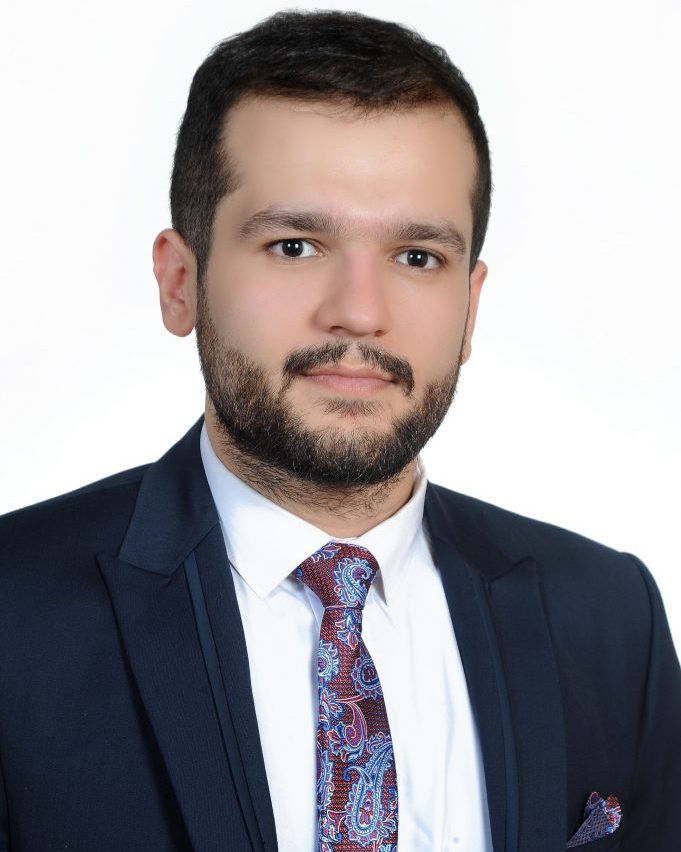}}]
{Reza Farahani} is a last-year Ph.D. candidate at the Institute of Information Technology (ITEC), Alpen Adria-Universität Klagenfurt (AAU). Currently, he is working on the Horizon Europe GraphMassivizer project. 
During the period from October 2019 to February 2023, he was involved in the ATHENA project, funded by the Christian Doppler Forschungsgesellschaft and the industry partner Bitmovin GmbH. From November 2022 to January 2023, he was a visiting scholar at the 5G \& 6G Innovation Centre (5GIC \& 6GIC), Institute for Communication Systems (ICS), at the University of Surrey, UK. He received his B.Sc. and M.Sc. degrees in computer engineering, in 2014 and 2019, from the University of Isfahan, Isfahan, IRAN, and the University of Tehran, Tehran, IRAN, respectively. He also has been working in the computer networks field in different roles, e.g., Cisco network engineer, network protocol designer, network programmer, and Cisco instructor (R\&S, SP) for over six years. His research interests are Multimedia Communication, Video Networking, Serverless Computing, Cloud-Edge Continuum, Network Softwarization/Virtualization, Distributed Systems, Mathematics Optimization, and Distributed Learning approaches. 
Further information at \url{https://www.rezafarahani.me/}.
\end{IEEEbiography}
\vspace{-1cm}
\begin{IEEEbiography} 
[{\includegraphics[width=1in,height=1.25in,clip,keepaspectratio]{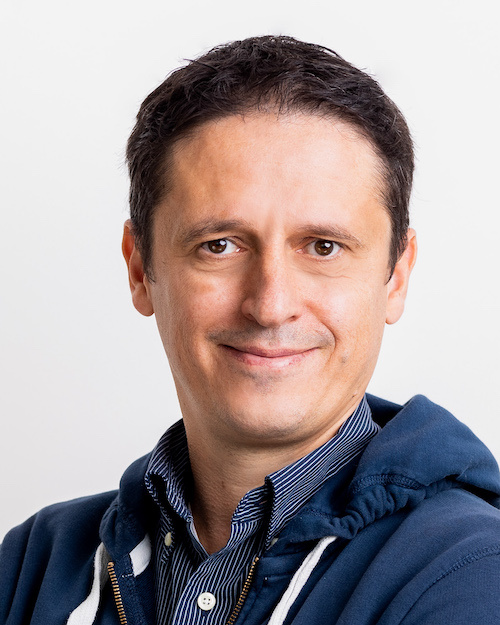}}]
{Christian Timmerer}\textbf{(M'08-SM'16)} is a full professor of computer science at Alpen-Adria-Universität Klagenfurt (AAU), Institute of Information Technology (ITEC), and he is the director of the Christian Doppler (CD) Laboratory ATHENA (\url{https://athena.itec.aau.at/}). His research interests include multimedia systems, immersive multimedia communication, streaming, adaptation, and quality of experience where he co-authored eight patents and more than 300 articles. He was the general chair of WIAMIS 2008, QoMEX 2013, MMSys 2016, and PV 2018 and has participated in several EC-funded projects, notably DANAE, ENTHRONE, P2P-Next, ALICANTE, SocialSensor, COST IC1003 QUALINET, ICoSOLE, and SPIRIT. He also participated in ISO/MPEG work for several years, notably in MPEG-21, MPEG-M, MPEG-V, and MPEG-DASH, where he served as standard editor. In 2012 he cofounded Bitmovin (\url{http://www.bitmovin.com/}) to provide professional services around MPEG-DASH, where he holds the position of the Chief Innovation Officer (CIO) -- Head of Research and Standardization. Further information at \url{http://timmerer.com}. 
\end{IEEEbiography}
\vspace{-.5cm}
\begin{IEEEbiography}
[{\includegraphics[width=1in,height=1.25in,clip,keepaspectratio]{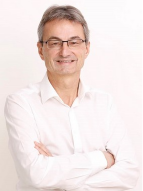}}]
{Hermann Hellwagner} 
is a full professor of computer science at Alpen-Adria-Universität Klagenfurt (AAU) where he leads the research group Multimedia Communication (MMC) in the Institute of Information Technology (ITEC). Earlier, he held positions as an associate professor at the University of Technology in Munich (TUM) and as a senior researcher at Siemens Corporate Research in Munich. His current research areas are distributed multimedia systems, multimedia communication and adaptation, QoS/QoE, and communication in multi-UAV networks. He has published widely on parallel computer architecture, parallel programming, and multimedia communication and adaptation. He is a senior member of the IEEE and of the ACM. He was a member of the Scientific Board of the Austrian Science Fund (FWF) (2005-2016) and an FWF Vice President (2013–2016). Currently, he is a member of the CD Senate of Christian Doppler Forschungsgesellschaft (CDG). Further information at \url{https://www.itec.aau.at/~hellwagn/}.
\end{IEEEbiography}

\end{document}